\documentclass[twocolumn,preprintnumbers,superscriptaddress,endnote,nofootinbib,prd]{revtex4}
\usepackage{graphicx}

\usepackage[hypertex]{hyperref}
\newcommand{\vev}[1]{\langle {#1} \rangle}
\newcommand{\lsim}{\lesssim}
\newcommand{\gsim}{\gtrsim}

\newcommand{\ord}[1]{\mathcal{O}{(#1)}}
\newcommand{\beq}{\begin{equation}}
\newcommand{\eeq}{\end{equation}}

\newcommand{\luv}{\Lambda_{\rm U\!V}}
\newcommand{\mP}{\bar{M}_{\rm P}}

\def\etal{{\it et al.}}

\begin{document}

% page numbers bottom-center
\pagestyle{plain}

\preprint{BNL-HET-08/3}
\preprint{YITP-SB-08-43}

\title{The Little Randall-Sundrum Model at the Large Hadron Collider}

\author{Hooman Davoudiasl
%\footnote{email: hooman@bnl.gov}
}
\affiliation{Department of Physics, Brookhaven National Laboratory,
Upton, NY 11973, USA}

\author{Gilad Perez
%\footnote{email: gilad.perez@stonybrook.edu}
}
\affiliation{C.N. Yang Institute for Theoretical Physics, State
University of New York, Stony Brook, NY 11794-3840, USA}

\author{Amarjit Soni
%\footnote{email: soni@bnl.gov}
}
\affiliation{Department of Physics, Brookhaven National Laboratory,
Upton, NY 11973, USA}

%%%%%%%%%%%%%%%%%%%%%%%%%%%%%%%%%%%%%%%%%%%%%%%%%%%%%%%%%%%%%%%%%%%%%%%%%%%%

\begin{abstract}

We present a predictive warped model
of flavor that is cut off at an ultraviolet scale $\ord{10^3}$~TeV.
This ``Little Randall-Sundrum (LRS)" model is a volume-truncation,
by a factor $y \approx 6$, of the RS scenario and is holographically
dual to dynamics with number of colors larger by $y$.  The
LRS couplings between Kaluza-Klein states and the Standard Model
fields, including the proton constituents, are explicitly
calculable without {\it ad hoc} assumptions. Assuming separate gauge
and flavor dynamics, a number of unwanted contributions to precision
electroweak, $Z b\bar b$  and flavor observables are suppressed in
the LRS framework, compared with the corresponding RS case. An
important consequence of the LRS truncation, independent of precise
details, is a significant enhancement of the clean (golden)
di-lepton LHC signals, by $\ord{y^3}$, due to a larger
``$\rho$-photon" mixing and a smaller inter-composite coupling.

\end{abstract}
\maketitle

%%%%%%%%%%%%%%%%%%%%%%%%%%%%%%%%%%%%%%%%%%%%%%%%%%%%%%%%%%%%%%%%%%%%%%%%%%%
Electroweak symmetry breaking (EWSB) in the Standard Model (SM)
via the Higgs condensate $v\equiv \vev{H}\simeq 250$~GeV
is economical and consistent with data.
However, quantum effects render $v$ quadratically sensitive to an
ultraviolet (UV) cutoff scale $\luv$.
For $\luv$ near the gravity scale
$\mP \sim 10^{18}$~GeV, a severe ``hierarchy"
$\ord{10^{-32}}$ arises.  One may
question the urgency of this problem, as physics
close to $\mP$ is unknown and inaccessible in the near future.
Nonetheless, precision
electroweak (EW) data require $\luv \gsim
10$~TeV,
near well-tested scales, posing
a challenge to a natural Higgs sector. This is often called
the little hierarchy; for some recent
proposals to address this problem see~\cite{ArkaniHamed:2001nc}.
Precision flavor data demand $\luv\gsim 10^2-10^3$~TeV, posing a
much more severe ``weak-flavor" hierarchy.

The Randall-Sundrum (RS) model \cite{Randall:1999ee} was originally
proposed to solve the hierarchy problem and yielded distinct
collider signatures \cite{Davoudiasl:1999jd}.  However, with
4D-sequestered fermions \cite{Davoudiasl:1999tf,Pomarol:1999ad},
tension with precision data generates a little
hierarchy \cite{Davoudiasl:1999tf,Csaki:2002gy}, the resolution of which
led to the inclusion of SM fermions
\cite{Grossman:1999ra} and gauge fields
\cite{Davoudiasl:1999tf,Pomarol:1999ad} in the 5D bulk.  This
also provided an attractive explanation of the SM flavor
structure \cite{Grossman:1999ra,Gherghetta:2000qt}, but made
the RS model less accessible to experiments
\cite{Davoudiasl:2000wi,KKgluon,moreKKgluon,Lillie:2007ve,
Djouadi:2007eg,Agashe:2007ki}. In
addition, the generic theory
requires more structure to be
consistent with oblique and
non-oblique precision tests \cite{Agashe:2003zs,Agashe:2006at}
and constraints from flavor changing neutral currents \cite{FPR}.
In what follows, ``RS" denotes the original hierarchy model
and all of its extensions.

While the RS construction has a compelling appeal,
as it allows a simultaneous resolution of the SM
hierarchy and flavor puzzles, it is premised on a
strong assumption.  That is, warping extends
over many orders of magnitude,
without any basic change in physics,
from the weak scale to the Planck scale.   Surely this assumption
needs to be put to an experimental test and we will discuss below
how this may indeed be possible, in a warped scenario with various
attractive features.

In this work, we use a volume-truncated RS background only to address
the hierarchy between the weak (IR) and flavor (UV) scales.
SM couplings to new physics, and hence the LHC phenomenology,
are explicitly set by the flavor structure without {\it ad hoc} assumptions.
The 5D UV scale $M_5$
is taken to be $\ord{10^3}$~TeV to suppress
light-flavor operators in this ``Little Randall-Sundrum (LRS)"
model \cite{otherlow}. We note that all EW and flavor data
are compatible with having $\luv \sim M_5$, where
additional physics may arise.

Keeping Yukawa dynamics unchanged by our
truncation, a number of unwanted contributions
to precision EW and flavor data are suppressed
within the LRS scenario, 
compared to the RS counterpart.  In fact, we will
show that any specific RS model is always more constrained than its
corresponding LRS counterpart.  
An exciting consequence of the LRS truncation is a much improved
prospect for discovery at the LHC, via clean di-lepton ``golden" modes,
since the couplings of the KK gauge bosons
to light fermions are enhanced while their couplings
to the heavy fields are suppressed.
Here and below we assume,
for simplicity, an IR-brane Higgs and tree level matching
for the gauge couplings, as discussed later.  Also,
we will focus on the quark sector,
however, leptons can be included straightforwardly.

The RS background is a slice of AdS$_5$, bounded by two Minkowski
3-branes, with the metric \cite{Randall:1999ee}
$
ds^2 = e^{-2
\sigma} \eta_{\mu\nu} dx^\mu dx^\nu - r_c^2 d\phi^2,
$
where $\sigma = k r_c |\phi|$, $k$ is the 5D curvature scale,
$r_c$ is the radius of compactification, and $\phi\in [0, \pi]$.
%and a $\mathbb Z_2$ orbifolding of the 5$^{\rm th}$ dimension is assumed.
The UV (Planck) brane is at $\phi=0$ and the IR
(TeV) brane is at $\phi=\pi$.  Going from the UV brane to the
IR brane, the 4D scale redshifts from $k\lsim M_5$ to the weak scale
$
\kappa\equiv k e^{-k r_c \pi}\sim~{\rm TeV}.
$
%AS5
%Resolving the 
Solution to the hierarchy problem requires $k r_c\simeq 11$, with
the Higgs at or near the IR brane.
A natural flavor structure is obtained, using
bulk fermions with non-zero vector-like masses
$m_i$, $i=u, d, \ldots$ \cite{Grossman:1999ra,Gherghetta:2000qt}.
The resulting zero-mode fermions are
exponentially localized in 5D,
parameterized by $c_i\equiv m_i/k$.  One may
choose $c_i\sim 1$ so that light fermions are UV-localized and
have small overlaps with the IR-localized Higgs.
Due to the warping, the light-flavor cutoff scale
is then much larger than the IR/weak scale.
This suppresses dangerous light-flavor operators and
yields the correct fermion mass hierarchy with $\ord{1}$ parameters.
However, not all precision data are accommodated with
bulk fermions.

%%%%%%%%%%%%%%%%%%%%%%%%%%%%%%%%%%%%%%%%%%%%%%%%%%%%%%%%%%%%%%%%%%%%
%%%%%%%%%%%%%%%%%%%%%%%%%%%%%%%%%%%%%%%%%%%%%%%%%%%%%%%%%%%%%%%%%%%%

{\it \underline{ Oblique Corrections}:}
Here we would like to go over the important constraints
on warped phenomenology from precision electroweak data.
There are various contributions that can be parameterized
in terms of the oblique Peskin-Takeuchi
$(S,T)$ parameters \cite{Peskin:1991sw} and we will discuss
them in turn.

We begin by considering the case where the only gauged symmetries 
in the bulk are those of the 5D SM.  First of all, there is 
a contribution that comes from the {\it tree-level} mixing of 
the gauge zero modes with the higher KK modes.  In the RS model, 
these are given by \cite{Agashe:2003zs},
\beq
S_{tree} \approx 2\pi\, (v/\kappa)^2\left[1 - \frac{1}{k r_c \pi} + \xi(c)\right],
\label{S2}
\eeq
\beq
T _{tree} \approx \frac{\pi}{2 \cos\theta_W^2}(v/\kappa)^2
\left[k r_c \pi - \frac{1}{k r_c \pi} + \xi(c)\right],
\label{T2}
\eeq
where
\beq
\xi(c)\equiv
\frac{(2c - 1)/(3 - 2c)}{1 - e^{k r_c \pi(2c-1)}}
\left(2k r_c \pi - \frac{5-2c}{3-2c}\right)
\label{xi}
\eeq
encodes fermion localization;
$\cos^2\theta_W\simeq 0.77$.  For all realistic warped 
fermion profiles of interest in this work, $\xi(c) \ll 1$.  

Without a ``bulk" custodial symmetry, there is also a UV-sensitive 
loop contribution $\delta T$ to the $T$ parameter.  
This UV-sensitivity can be
absorbed into a higher dimension operator.  Assuming that 
this operator is generated by strong dynamics at the 4D cutoff
scale $\Lambda$, it will have the form
\beq
16 \pi^2\frac{(D^\mu H)^\dagger H(H^\dagger D_\mu H)}{\Lambda^2}.
\label{dim6}
\eeq

Current data favor
$|S| \sim |T| \sim 0.1-0.3$ \cite{Yao:2006px} summed 
over all contributions.  We see that $T_{tree}$ from 
Eq.~(\ref{T2}) is the dominant tree-level 
constraint, given its volume enhancement 
$k r_c \pi$, which is roughly a factor of 35 in the RS model.  
To reduce the size of $T_{tree}$ in this setup 
then requires increasing
$\kappa$ ($m_{KK}$) to values that lead to
a severe little hierarchy and null LHC signals.  
Alternatively, we see that reducing $k r_c \pi$ yields a
significant suppression, keeping KK masses fixed.  In our 
LRS construct, we will truncate the volume to $k r_c \pi= 6$.  
Note that $m_{KK}= x_{KK}\, \kappa$, where 
$x_{KK}^{RS} = 2.45, 5.56, \ldots$ 
\cite{Davoudiasl:1999tf,Pomarol:1999ad} and 
$x_{KK}^{LRS} = 2.70, 5.87, \ldots$.  Then, 
for $m_{KK} \approx 5$~TeV, the RS model yields 
$(S,T)_{tree} \approx (0.1, 1.1)$, whereas for the 
LRS model $(S,T)_{tree} \approx (0.1, 0.2)$, 
from Eqs.~(\ref{S2}) and (\ref{T2}).

Even though the LRS truncation has 
suppressed the tree-level KK-tower mixing 
contribution, we must still address the loop and 
cutoff scale effects encoded in Eq.~(\ref{dim6}).  These are the same 
in both the RS and the LRS scenarios, since the 
flavor sector is assumed to be independent of the 
gauge dynamics and hence unchanged by the LRS truncation.  
Barring unnatural cancellations, demanding
that the cutoff contribution is less than $\ord{0.1}$
pushes $m_{KK}$ to values of $\ord{10}$~TeV. 

Ref.~\cite{Agashe:2003zs} attributed $T\gsim 1$, in the RS model, 
to the absence of bulk ``custodial" protection and postulated
a $SU(2)_L \times SU(2)_R \times U(1)_X$ 5D symmetry
to eliminate tree-level
contributions to $T$.  It turns out that the loop contribution
$\delta T$, governed by fermion KK modes, still remains, but is 
no longer UV-sensitive.  Also, given the gauged 5D 
custodial symmetry, there is no cutoff contribution of the form 
in Eq.~(\ref{dim6}), at the IR-boundary.  Ref.~\cite{Agashe:2003zs}
concluded that including the SM effects, for
$m_{KK}^{RS} \sim 3-4$~TeV and a light Higgs,
$S$ and $T$ can be accommodated at an acceptable level.  
Note that with a bulk custodial symmetry, the LRS construct
will enjoy the same level of agreement with the 
oblique data as the models in Ref.~\cite{Agashe:2003zs}.

The suppression of $T_{tree}$ in the LRS scenario, without bulk
custodial symmetry, can be understood as follows.  After
EWSB, the entire KK tower of states mixes
and the zero mode (SM)
wavefunctions get deformed away from a constant. This generates a
large tree-level oblique correction to $T$, controlled by the KK-Higgs ({\it i.e.}
KK-IR brane) coupling.  The deformation of the
SM $W/Z$ wavefunction is then proportional to $k r_c
\pi$.  Hence, the LRS contributions get suppressed by a factor
\beq
y \equiv
\frac{k r_c \pi|_{RS}}{k r_c \pi|_{LRS}}\approx 6.
\label{y}
\eeq

Note that $S$ from Eq.~(\ref{S2}) is basically the same in the RS and LRS models.  
This is because the dominant contributions to $S$ come from a universal shift in
the light fermion-gauge field couplings~\cite{Agashe:2003zs}.
This shift depends on the product of the mixing between the zero modes
and KK gauge states (after EWSB) and the universal couplings of KK gauge fields
to the light fermions. While the former decreases
with shrinking volume the latter is increased
so that the product is unchanged.

In brief, the LRS truncation 
does suppress the contribution from KK-tower mixing 
in the gauge sector, compared to the RS case, 
quite efficiently.  However, a bulk custodial 
symmetry is still required in both the RS and LRS 
setups to control loop and cutoff scale contributions to $T$ and 
to bring the scale of KK masses down to 
$\sim 3$~TeV~\cite{Agashe:2003zs,Carena:2007ua}.   
As we will show next, a more dramatic
improvement can be achieved regarding the non-oblique and precision
flavor observables.

%%%%%%%%%%%%%%%%%%%%%%%%%%%%%%%%%%%%%%%%%%%%%%%%%%%%%%%%%%%%%%%%%%%
{\it \underline{Non-oblique Correction \& Flavor Physics}:}
Refs.~\cite{aps,NMFV} have shown that RS
models flow to next-to-minimal-flavor violation, where
flavor changing effects are primarily from mixing with the third generation.
The extra flavor breaking sources are quasi-aligned with their SM counterparts
and the misalignment is
at most of order the CKM matrix, but new sources of CP violation are present.
Non-oblique $Z b {\bar b}$ and FCNC constraints are more involved since
they depend on the amount of flavor non-universality,
determined by the fermion zero-mode IR-brane profile values, $f_{Q,u,d}$~\cite{aps}.

As shown in Ref.~\cite{aps,hs}, once the overall scale of
5D Yukawa coupling $\lambda_5$ and $t_R$-localization $c$-parameter
are fixed, the localization of all other
fermions is set by the measured masses and CKM
mixing angles, assuming anarchical 5D Yukawa
matrices via the relation $m_{u,d}\propto F_Q \lambda_5^{u,d} F_{u,d}$,
where $m_{x}$ denotes 4D masses and $x=u,d$
correspond to the up and down quarks respectively;
$f_x$ are eigenvalues of $F_x$.  Here, we will keep $\lambda_5$ unscaled
by LRS-truncation and fixed at
its ``RS value"~\cite{aps,FPR} (see discussion below).  Then, the amount of non-universality is
unchanged, but the strength of the KK-mediated effects get decreased like
the truncated LRS volume, parameterized by $k r_c \pi$.
This is the reason the non-universal precision observables
are significantly suppressed in
our model.  For concreteness, in Table~I, we give
a set of $c_x$ and $f_x$ values
that reproduce the SM quark masses and mixing angles; $\lambda_5/k = 2$
in accordance with Ref.~\cite{aps}.
\begin{table}[!hbt]\begin{center}
\begin{tabular}{||c|l|l|l||}
    \hline\hline
    { Flavor}& { $c_Q,\,f_Q$} & { $c_u,\,f_u$} & { $c_d,\,f_d$}\cr
    \hline\hline
    I & 1.45,\,0.003& 1.7,\, $8\times10^{-4}$ &1.52,\,0.002\cr \hline
    II& 1.17,\,0.015& 0.86,\,0.071 &1.26,\,0.009\cr\hline
    III &0.52,\, 0.28& $- 0.19$,\,0.83 &1.14,\,0.018\cr
\hline\hline
 \end{tabular}

\caption{{\small The eigenvalues of $c_x$ and $f_x$ which roughly yield the
right masses and CKM elements at the $m_Z$ scale~\cite{PDG}.}}
\end{center}\end{table}

Bulk-RS constraints from the $Z b {\bar b}$ coupling
require non standard fermion representations under
the custodial symmetry, as well as
a $Z_2$ symmetry \cite{Agashe:2006at} ,
in order to have
$m_{KK}^{RS} \sim 3$~TeV; otherwise,
$m_{KK}^{RS} \gsim 5$~TeV \cite{Agashe:2003zs}.
Without the custodial symmetry, there are various
contributions to $Z b\bar b$.
The first originate from the gauge zero-mode-KK-tower mixing due to EWSB and the
enhanced coupling of the gauge KK modes to IR-localized $b_L$.
These corrections are proportional
to $(k r_c \pi/m_{KK}^2) f^2_{Q^3}$ \cite{Agashe:2003zs}, where
$f_{Q^3}$ is assumed to have the RS value.
Thus, the LRS bound from these contributions is
\beq
m_{KK}^{LRS} \gsim m_{KK}^{RS}/\sqrt{y}. \hskip 1cm (Z b {\bar b})
\label{mkklrs}
\eeq
With
$\sqrt y\approx 2.4$ we get
$m_{KK}^{LRS} \gsim 2$~TeV.

The second type of correction to $Z b\bar b$ is due
to $\ord{1}$ mixing between $b_L$ and the exotic $SU(2)_R$
partner of $t_R$ \cite{KA}.  This latter contribution will be absent for
a choice of representation in which $t_R$ is a $L$-$R$
isosinglet \cite{Agashe:2006at}.  Note that without a bulk custodial symmetry,
there is no exotic
$t_R$ partner. However, there is a third type of correction to $Z
b\bar b$ from the mixing of the KK modes of $b_R$ and the $b_L$ zero
mode.  This contribution is not truncated in the LRS model and is of
order $4 [(v/\sqrt{2})\lambda_5 k f_{Q^3}/m_{KK}(b_R)]^2$.  To keep
deviations in the $Z b\bar b$ coupling below $0.3\%$, we then need
$m_{KK}(b_R)\gsim 4$~TeV.  Interestingly, $m_{KK}\gsim 3$~TeV
for gauge fields already implies the former bound for the KK modes of
$b_R$, in the LRS framework presented here.  We hence conclude that
all of the above constraints from $Z b\bar b$ can be satisfied for
gauge sector $m_{KK}\gsim 3$~TeV, without any
protective symmetries (however, as discussed above,
this would be inconsistent with the bound from the $T$ parameter).

%an acceptable value for $T$ would then require $m_{KK}^{LRS}\gsim 5$~TeV,
%as discussed before.

We finally review the strongest constraints on generic bulk RS models,
from $\Delta F=2$  processes due to tree level exchange of KK gluons.
Ref.~\cite{aps} showed that, with an IR localized Higgs, the ratio
of RS and SM  $(V-A)\times (V-A)$
contributions $h^{\rm RS}\propto (F_Q^2)_{ij}^2$
(in the down quark mass basis).
One can write
\beq
h^{\rm RS}={{M_{12}^{\rm RS}}\over M_{12}^{\rm SM}}
     \sim{0.5}\times \frac{k r_c \pi}{35} \left({3 {\rm TeV}\over m_{\rm
           KK}}\right)^2
     \left({{ f_{Q^3} \over 0.3} }\right)^4\,.
\eeq
At present, $h^{\rm RS}\lsim 0.3$~\cite{ENP,NMFV,problem}. However,
the dominant contribution $\delta(\epsilon_K)$
to $\epsilon_K$ from
$(V-A)\times (V+A)$ operators
~\cite{problem} is given by $\delta(\epsilon_K)\propto
k r_c \pi  (F_Q^2)_{12} (F_d^2)_{12}$ \cite{FPR} which is
${\cal O} (20)$ times smaller. This is
not enough due to a matrix element
chiral enhancement of ${\cal O}(11)$ and a
${\cal O}(7)$ factor from the running between the KK
and weak scales, requiring $m_{KK}^{RS}\gsim 8$~TeV.
In the LRS case, both contributions are
suppressed by $y$ and thus $m_{KK}^{LRS}\gsim 3$~TeV,
no stronger than the oblique constraints.
We note that the RS CP electric dipole moment problem \cite{aps}
persists in our setup, as it is governed by
5D Yukawa interactions which are unchanged
(for possible solutions to this problem see \cite{FPR,CFR}).

%%%%%%%%%%%%%%%%%%%%%%%%%%%%%%%%%%%%%%%%%%%%%%%%%%%%%%%%%%%%%%%%%%%%%%

{\it \underline{Phenomenology}:}
Gauge KK modes couple to UV-localized light fermions (important initial
states at colliders),
with strength $g_{KK}\sim g_4/\sqrt{k r_c \pi}$; $g_4$ is a
typical 4D SM gauge coupling.  We get $g_{KK}^{RS} \sim g_4/6$ and
$g_{KK}^{LRS}\sim g_4/2.5$.  In particular,
the UV-brane values of the normalized first gauge KK
wavefunctions $\chi^{(1)}$ are
\beq
\chi^{(1)}_{RS}(\phi=0) = -0.08 \;\; ; \;\; \chi^{(1)}_{LRS}(\phi=0) = -0.20\,.
\label{chiRS}
\eeq
This leads to improvements in the LHC sensitivity, at fixed $m_{KK}$,
for the following reasons:
{\it (i)} Typically broad states~\cite{KKgluon} become narrower by
a factor $y\sim (0.2/0.08)^2$,
{\it (ii)} branching ratio (BR) into light states ({\it e.g.} $e^+ e^-$)
increases by a factor $y^2$, {\it (iii)} from {\it (i)} and
{\it (ii)} one can show that the signal ${\cal S}$ goes up by $y^3 \sim 250$,
while the background ${\cal B}$ drops as $1/y$, over the resonance width.
Hence, ${\cal S}/{\cal B}$ in the LRS model
is expected to go up by a factor $y^4 \sim 1500$, a remarkable enhancement!
These features lead to
a larger LRS discovery reach and a way to test the validity of this
setup and the underlying assumptions.
As the LHC reach for KK gluons in bulk-flavor RS
models is 3-4~TeV \cite{KKgluon}, the corresponding LRS reach can be
as big as $\sim 5$~TeV.  The enhanced $g_{KK}$ in the LRS model
could also allow access
to the elusive EW gauge KK modes~\cite{Agashe:2007ki}.
For example, the $Z'\to \ell^+ \ell^-$, $\ell=e,\mu$,
golden decay modes which were close to
hopeless within the RS case~\cite{Agashe:2007ki}
could lead to discovery in the LRS setup. Using the same
cuts as in Ref.~\cite{Agashe:2007ki},
we find roughly 2000 (3) events with
${\cal S}/{\cal B}$  and ${\cal S}/\sqrt {\cal B} \gg 1$ [in fact $\ord{100}$]
for $M_{Z^\prime} =2\ (5)~$TeV and 100~fb$^{-1}$ \cite{Piai}.
Note that given the significance of the signal, a discovery would be
unambiguous over this range.
Enhancement in production rate is expected for the SM KK
fermions whose LHC discovery, in the RS model, would be quite
challenging \cite{Davoudiasl:2007wf}.
Furthermore, any
bulk couplings mediated via $\lambda_5$
are relatively stronger due to LRS scaling. This
would, in principle, be a direct test of our setup.
Also, BR's of the neutral modes into composite
states such as $W_L W_L,Z_L h$ and $t \bar t$ compared
to those into light fermions
will provide a robust test of the LRS construct.

TeV-scale spin-2 ``graviton" resonances are distinct RS signatures
\cite{Davoudiasl:1999jd,Davoudiasl:2000wi}.  Since $M_5 \sim
10^3$~TeV, we eliminate the zero mode graviton, using UV-brane
Dirichlet boundary conditions; we find $x^{(LRS)}_G \simeq
x^{(RS)}_G = 3.83, 7.02,\ldots$. Gluon-fusion production of KK
gravitons is dominant at hadron colliders, with a cross section
proportional to $[(k/M_5)/k r_c \pi]^2(k/M_5)(x_G/m_G)^2$
\cite{Davoudiasl:2000wi}. In a generic LRS model, both $k/M_5$ and
$k r_c \pi$ shrink by a factor of $y$, decreasing this cross section
by a factor of $\ord{y}$.  Thus, observation of the LRS
spin-2 resonances is unlikely at the LHC (probably more unlikely 
than in the RS case \cite{Davoudiasl:2000wi}).

We also note that there can be LRS collider signatures in 
terms of deviations 
from the SM top-quark couplings.  Within the RS scenario, there are two types
of contributions to $t\to cZ$ of similar size \cite{TopFCNC}.
One is through mixing between $Z$ zero and KK modes
which will be suppressed by $y$ and probably
unobservable in the LRS framework. However, the second one proceeds through
mixing between $t_R$ and the KK modes of $t_L$.
This mixing is controlled by the 5D Yukawa which
is left unchanged in our LRS construct and, therefore,
$t\to cZ$ should be within the reach of the LHC.
A similar effect yields an $\ord{20\%}$ shift in the
$Z t_R \bar t_R$ coupling which is probably beyond the LHC sensitivity but
may be observed at a future linear collider \cite{moretcZ}.

\begin{table}[!hbt]\begin{center}
 \begin{tabular}{||c|c|c||l}
    \hline\hline
    constraint/prediction & RS & LRS\cr
    \hline\hline
    $T$ parameter & 3& 3  \cr \hline
    $S$ parameter &3 & 3 \cr\hline
    $Z\to b\bar b$ &3 & 3$^*$ \cr\hline
    $\epsilon_K$  &8 & 3 \cr\hline
    ${\cal S}/{\cal B}$ for $Z'\to l^+ l^-$ &\{0.3, $-$\} & 
    \{$\ord{100}$, $\ord{100}$\} \cr
\hline\hline
 \end{tabular}

\caption{{\small Summarized comparison of constraints and predictions  
in the RS and the LRS scenarios. 
For simplicity and definiteness, the Higgs is assumed to be on the IR-brane.
The constraints correspond to lower bounds on gauge KK masses, in TeV. 
Here, we assume a custodial symmetry for the $T$ parameter;  
a left-right $Z_2$ symmetry is imposed to protect 
the $Z b\bar b$ coupling, unless denoted by $^*$.  The predictions in the 
last row correspond to a $Z^\prime$ of mass \{2, 5\}~TeV, respectively.
}}
\end{center}\end{table}

%%%%%%%%%%%%%%%%%%%%%%%%%%%%%%%%%%%%%%%%%%%%%%%%%%%%%%%%%%%%%%%%%%%%%%

{\it \underline{Holography}:} We now present a qualitative
discussion of the LRS, using the AdS/CFT correspondence
\cite{Maldacena:1997re}, following previous interpretations of
geometric RS results \cite{ArkaniHamed:2000ds,Rattazzi:2000hs}
in the dual context of a strongly coupled large $N$
4D gauge theory \cite{'t Hooft:1973jz}.
We begin by studying the effects of LRS-truncation on
the classical geometric relation
between $g_4$ and the 5D gauge
coupling $g_5$ \cite{Agashe:2002bx}:
\beq
1/g_4^2 = \tau_{\rm
UV} + \tau_{\rm IR} + \log(k/\kappa)/(k g_5^2).
\label{g42}
\eeq
Here, $\tau_{\rm UV}$ and $\tau_{\rm IR}$ will be treated as small
UV and IR quantum threshold corrections, respectively.  For a
generic comparison of couplings, we neglect
$\tau_{{\rm UV},{\rm IR}}$ and keep $g_4$ fixed to its measured value.
Thus, reducing the volume suppression $k r_c \pi$ (the $\log$)
requires lowering the value of $k g_5^2$.  In the dual CFT,
this is interpreted as the contribution of
CFT ``quarks" to the running of external gauge couplings from the
fundamental scale, $M_5$, down to the TeV scale (just like
the contribution of SM quarks to $\alpha_{QED}$ running)
\cite{ArkaniHamed:2000ds,Rattazzi:2000hs,Agashe:2002jx}. Thus the
relation $\sqrt{k g_5^2}\sim 4 \pi/\sqrt{N}$ should hold
between the dual theories. Explicit calculations
\cite{Davoudiasl:1999tf,Pomarol:1999ad}
confirm that couplings among gauge KK modes, {\it i.e.}
IR localized fields,
are enhanced, compared to the corresponding zero mode gauge coupling, 
by $\sqrt{kr_c \pi} \sim \sqrt{ k g_5^2} $.  This, in the dual CFT picture, 
corresponds to the coupling of
three composites given by $4 \pi/\sqrt{N}$ at large $N$
\cite{'t Hooft:1973jz,Witten:1979kh}. Consequently, the truncated
LRS volume is dual to $N^{LRS}\sim y N^{RS}>N^{RS}$, making the
inter-composite interactions weaker.  The weakened CFT interactions
with the Higgs, a composite state, account for the decrease in $T$ from
Eq.~(\ref{T2}).

In our LRS construct, we held the 5D Yukawa coupling $\lambda_5$
unscaled, lowering the 4D IR-brane
cutoff to about 10~TeV, as
in the RS case.  In the dual language, this corresponds to
separate dynamics for this sector, characterized by a ``flavor" CFT
with $N_F\sim N^{RS}<N^{LRS}$ ($N_F\sim 3-4$ \cite{aps,FPR}).  This
independent CFT is linked to dynamical
breaking of the 5D flavor group which is not completely broken by
the bulk masses (unlike $\vev{H}$ which breaks the ``chiral"
symmetry) \cite{aps}.  If the Higgs is
dynamically realized as a pseudo-Goldstone boson (PGB) a similar
different scaling for its potential should be applied, {\it i.e.}
the dynamics which generates the PGB potential is
characterized by $N=N_F$. Otherwise, increasing  $N$ would
induce a more severe fine tuning for the PGB potential \cite{Zbbar}.
However, in all the known models including the most
realistic ones \cite{Zbbar} the dominant contributions to this
potential come from the top sector, corresponding to the flavor CFT,
and not from the weakly gauged
one, consistent with the above assumptions~\cite{CHG}\footnote{However,
for the particular minimal realization of a PGB Higgs as an $A_5$, flavor and
gauge dynamics are of the same origin which implies only a
single value of $N$.}.  This also explains why
keeping $S \sim (v/f_\pi)^2 N^{LRS}$ at the RS value does not lead to extra
fine-tuning of $v$, since the ``decay constant"
$f_\pi$ here is from the weakly gauged dynamics
which does not govern the Higgs potential.
The constancy of the $S$ parameter under truncation
can be understood as follows.
The main contribution
to $S$ is from the universal vertex corrections~\cite{Agashe:2003zs}.  This is
controlled by gauge
zero-KK mode mixing, which scales as $1/\sqrt{N}$,
and the universal KK couplings to light
fermions, which scales as $\sqrt{N}$ (see below). Therefore, $S$
remains unchanged.

The non-oblique and FCNC contributions depend on the amount of
non-universality in the couplings of the KK states to different
generations. On the CFT side, this corresponds to the amount of
partial compositeness for a given $N_F$.  The
amount of compositeness follows from the observed masses and mixing
angles \cite{aps}, once $\lambda_5$ is
set and the location of $t_R$ is decided.
By fixing these to the RS value,
the LRS amount of partial compositeness is then unchanged, and hence
the non-universal observables are suppressed by truncation.
This, generically, yields a better agreement with the
data. It implies that interactions proportional to $\lambda_5$
(such as between  the Higgs and two KK fermions) are
stronger than the corresponding KK gauge interactions.

An important consequence of volume truncation is
enhanced $\rho-photon$ mixing, proportional to $\sqrt N$
\cite{ArkaniHamed:2000ds,Agashe:2002jx}, leading to
larger couplings of light SM fermions
to gauge composite/KK modes.  The composite (KK) partial widths
into elementary fermions scales as $N$, while the
total width drops as $1/N$. Hence, ${\cal S}\sim N^3$ and ${\cal B}\sim 1/N$,
over the resonance width. Both effects yield stronger LRS signals
at the LHC than for the RS case, since $y = N^{LRS}/N^{RS} \gg 1$, as discussed before.
This is analogous to how $e^+ e^-\to \rho \to \mu^+\mu^-$ is modified
when $N_c$ is increased.

Finally, we emphasize that unless mentioned
explicitly (as for $\lambda_5$) we have rescaled all
the couplings in the theory according to the LRS value of $N$.
This is why we did not get an enhancement in the KK
graviton production.  Also,
we have neglected brane-kinetic terms to allow
a transparent comparison
of our model with {\it generic} RS models
where such terms have
sub-dominant effects on
the above observables.  Lastly, we note that
$M_5\sim 10^3$~TeV does not suppress baryon and lepton
number violation sufficiently.
Such issues lie beyond the scope of this work, but
can be addressed with discrete symmetries or in a
UV-completion of the LRS model.
However, dimension-9 operators suppressed by the LRS $M_5$
lead to $n-\bar n$ oscillations at acceptable levels \cite{BaldoCeolin:1994jz}
and may be accessible in near future
experiments.

%%%%%%%%%%%%%%%%%%%%%%%%%%%%%%%%%%%%%%%%%%%%%%%%

In summary, we presented the
``Little Randall-Sundrum (LRS)" model  of
hierarchy between the flavor and weak scales which is much
less constrained than $\mP$-weak warped scenarios.  Here,
the 5D cutoff scale  $M_5 \sim 10^3$~TeV is chosen to suppress unwanted
light-quark operators sufficiently and the weak
scale is obtained from $\ord{M_5}$ scales by warping; the
flavor puzzle is addressed by fermion localization, as in the RS
model.  Even without a bulk custodial symmetry, the
``tree-level" lower bound on
LRS gauge KK masses is at $\sim 5$~TeV; the RS bound
is $m_{KK}^{RS}\gsim 12$~TeV.  Loop and higher dimension
contributions to $T$ raise $m_{KK}\gsim \ord{10}$~TeV, without
a custodial symmetry, in both RS and LRS models.
Custodial symmetry can
be imposed if desired, leading to an oblique lower bound
at $\sim 3$~TeV. As we kept the overall Yukawa scale unchanged,
the most severe RS-type constraints are much better
behaved here: non-oblique constraints from $Z b {\bar b}$, without
a protective $Z_2$ symmetry, are absent for
$m_{KK} \gsim 3$~TeV and the FCNC bounds are largely relaxed.
We have summarized the comparison between the RS and LRS
frameworks regarding various precision constraints in Table~II.
As can be seen from this table, the LRS framework is never
more constrained than its RS counterpart and in many cases it is
much more compatible with data.
Typical LRS graviton KK modes are more elusive than those
of $\mP$-weak warped models.  However, the light-fermion
LHC-production rate and BR's
for LRS gauge
KK modes are much
bigger than the corresponding RS values and yield a
signal $\sim y^3$ times larger; $y\sim 6$ is the LRS
truncation factor.
We hence conclude that the LRS model is a
good candidate for new physics and may soon
be uncovered at the LHC, or perhaps probed at a future linear collider.

%%%%%%%%%%%%%%%%%%%%%%%%%%%%%%%%%%%%%%%%%%%%%%%%%%%%%%%%%%%%%%%%%%%

We thank Csaba Csaki, Shri Gopalakrishna,
Ami Katz, and Rob Pisarski for useful discussions.
We especially thank Kaustubh Agashe for many helpful comments.
G.P. thanks the Aspen Center for Physics and
the HEP groups at Boston and Harvard universities for hospitality
while working on this paper.
The work of A.S. and H.D. is supported in part by the United States Department of
Energy under Grant Contracts DE-AC02-98CH10886 and the work of G.P. is supported by the NSF
under grant 06353354.

%%%%%%%%%%%%%%%%%%%%%%%%%%%%%%%%%%%%%%%%%%%%%%%%%%%%%%%%%%%%%%%%%%%%%%%%%%%


\begin{thebibliography}{99}
\bibitem{Piai}
Similar LHC reaches were found, in a different context, by M. Piai in Ref.~\cite{otherlow}.
\end{thebibliography}
\end{document}